\documentclass[aps,prl,preprint,]{revtex4}

\def\ben{\begin{enumerate}} \def\een{\end{enumerate}}
\def\beq{\begin{equation}} \def\eeq{\end{equation}}
\def\bea{\begin{eqnarray}} \def\eea{\end{eqnarray}}
\def\beann{\begin{eqnarray*}} \def\eeann{\end{eqnarray*}}
\def\beasn{\begin{sneqnarray}} \def\eeasn{\end{sneqnarray}}
\begin{document}

\title{Reparameterization invariants for  anisotropic Bianchi I cosmology with a massless scalar source}
\author{D. C. Salisbury, J. Helpert, and A. Schmitz}
\date{November 2, 2007}

\affiliation{Department of Physics,
Austin College, Sherman, Texas 75090-4440, USA\\E-mail:
dsalisbury@austincollege.edu}

\begin{abstract}
Intrinsic time-dependent invariants are constructed for classical, flat, homogeneous, anisotropic cosmology with a massless scalar material source. Invariance under the time reparameterization-induced canonical symmetry group is displayed explicitly.

\end{abstract}


\maketitle

\section{Introduction}
\label{intro}

There is now general agreement in the relativity research community that time is a relational concept, yet disagreements persist both on the context in which time is "recovered" and the role, if any, of the general covariance of Einstein's classical theory of general relativity. The contextual stances range all the way from the now largely dismissed claim that no true classical observable can vary with time (``frozen time''), to the expectation that variation in time will emerge only for appropriately selected quantum states from a timeless quantum theory of gravity. Attitudes regarding the underlying symmetry range from assertions that the full diffeomorphism symmetry group is lost forever in initial value formulations of general relativity, to our counterclaim that the full symmetry group is alive and well in canonical gravity. Indeed, our intention in this paper is to show how this canonical symmetry group can be used to introduce an observable relational time in a non-trivial gravitational model, an anisotropic universe with a massless scalar source. We will employ a canonically fixed intrinsic time, and we will compare and contrast our symmetry-based use of this time with several other approaches discussed in the literature.

It is surprising that although it has long been advocated that intrinsically defined time may be employed in general relativity some researchers still believe that quantities that are invariant under the action of the full four-dimensional diffeomorphism group in canonical general relativity are constant in time, i.e., time is ``frozen".   We believe this confusion persists in part because the full nature of the canonical symmetry group of general relativity has only recently been understood  \cite{ps04}. Crucial to this understanding is a distinction between gauge symmetry transformations and time evolution. Thus it is possible for the Hamiltonian to commute with invariants, which themselves commute with the individual constraints with which the gauge symmetry generator is constructed,  and yet these invariants are not constants of the motion.

The first proposal that material sources can be employed to intrinsically identify spacetime events goes all the way back to Einstein as he first rejected and then ultimately embraced the principle of general covariance in his general theory of relativity, as discussed in a classic paper by Stachel   \cite{stachel89}. The suggestion to use a spacetime scalar function of purely gravitational dynamical variables as an intrinsic clock was first made by G\'{e}h\'{e}niau and Debever  \cite{gd56a,gd56b}. Komar and Bergmann showed subsequently that Weyl scalars could be expressed in terms of canonical phase space variables and utilized to fix intrinsic coordinates in a classical Hamiltonian formulation of general relativity \cite{komar58,bergkom60,berg61a,berg61b}.   Dewitt was an early advocate for the use of intrinsic coordinates \cite{dewitt62}. Rovelli began to argue forcefully for their limited use in quantum gravity in the late 1980's \cite{rovelli91a,rovelli91b,rovelli91c}. Rovelli proposes a construction in which  parameter(coordinate) and internal (intrinsic) time will in general not coincide. But the fixing of a parameter time in terms of a canonical intrinsic time is indeed the fundamental idea of this paper. Indeed, as we shall show, since our intrinsic time can be achieved through a time parameter transformation, every dynamical variable evaluated in the chosen special coordinate system is an invariant under time reparamterizations, and hence an observable. 

Marolf was the first to seriously apply the idea of intrinsic time to quantum cosmology \cite{marolf95} .   (See the review articles by Isham and Kuchar, and also a more recent paper by Smolin for the conventional analysis of the problem of time, including arguments for and against the generic existence of intrinsic time  \cite{isham92,kuchar92,smolin00}.  

It turns out that our cosmological model is exactly solvable with an arbitrary choice of time coordinate.  We are thus able to demonstrate in detail that our phase space invariants do not change when the coordinate time is altered under the action of the canonical diffeomorphism-induced symmetry group presented by J. Pons, D.C. Salisbury and L. C. Shepley 
\cite{pss:1997pr,pss:2000jmp,pss:2000grg,pss:2000pr}. We construct invariants by choosing the evolving value of one of the cosmological expansion factors as an intrinsic time, i.e., by establishing  correlations between this value and the scalar field plus the spacetime metric components, including the lapse function. As explained in detail in  \cite{ps04}, the selection of an intrinsic time coordinate is equivalent to the establishment of a time-dependent gauge condition expressed in terms of phase space variables. Invariants under the action of the time-reparameterization-induced symmetry group discussed in the series of papers cited above can equivalently be constructed through the finite action of this group along a gauge orbit. The lapse function in our cosmological model is one of the bases of this gauge transformation group. This is true since all coordinate time reparameterizations have a compulsory dependence on the lapse if we wish for them to faithfully reproduce in phase space the time-reparameterization symmetries present in configuration-velocity space. The lapse must be intrinsically fixed, otherwise the gauge freedom is restricted as we shall show explicitly for the dynamical-variable-independent choice for the lapse made by Marolf \cite{marolf95}. There are three essentially new features in our construction of invariants in cosmology. First, we carefully distinguish between reparameterization symmetry (that requires the retention of the lapse as a dynamical variable) and coordinate time evolution. Second, we canonically generate infinitesimal variations of our putative invariant phase space functions and show that these variations vanish.  Along the way we will reveal in this non-trivial model how invariants can have vanishing Poisson brackets with the Hamiltonian, but can still have a non-trivial coordinate time dependence. And finally, all of our invariants satisfy the original second order Einstein equations.

In section 1 we give a brief overview of the general canonical symmetry group. We shall assume that  coordinates have been chosen for our homogeneous cosmology such that all fields are independent of the spatial coordinates. Thus the only interesting diffeomorphism symmetry remaining is reparameterization of the coordinate time.  We show in section 3 that our coordinate choice may be made in the Einstein action, yielding a finite dimensional Hamiltonian model. We find the general solution of the equations of motion in section 4. In section 5 we discuss the nature of the reparameterization-induced canonical symmetry group, and construct its generator.  Then we construct variables that are invariant under the action of this generator. We conclude the section with a counting of the degrees of freedom of the model. 

The sourceless (Kasner) model has been quantized by Bojowald in the context of loop quantum gravity, as has the isotropic model with massless scalar source \cite{bojowald03a,bojowald02}. Our analysis will be exclusively classical.  Some loop quantum gravitational implications are addressed elsewhere.

\section{The diffeomorphism-induced  transformation symmetry group and intrinsic coordinates}

We briefly summarize here the results of Pons, Salisbury and Shepley, and Pons and Salisbury, henceforth identified respectively as (I) and (II).  A group of transformations is said to constitute a symmetry if it leaves the form of the Lagrangian equations of motion unchanged.  One consequence of this definition is that in order to be able to compare the form of the second order Einstein equations that follow from the Hamiltonian formulation of general relativity, we need to know how to transform both the dynamical variables and the coordinates under the action of the full diffeomorphism group. This means that we must be continually cognizant of the relation between symmetry variations of the dynamical variables and the associated diffeomorphisms. Of course, another consequence of this definition is that under the symmetry transformation the full set of solutions to the second order equations of motion is transformed into itself. 

It has been standard practice, initiated by Dirac, to eliminate lapse and shift from the start as canonical variables. An immediate, obvious, and fateful outcome of this decision is that generically only the spatial diffeomorphism group survives as a symmetry. Obviously, with a fixed choice for time-dependent lapse and shift one is committed to a fixed time foliation of spacetime. And furthermore, it is apparent that the set of solutions (for this fixed choice of lapse and shift) does not include the complete set of solutions of the second-order equations of motion. The crucial observation here is that symmetries must transform the complete set of second order solutions into itself. If one insists in general relativity for example in keeping only the spatial metric and associated momentum as canonical variables, then one confronts the problem of how to construct a Hamiltonian in which the lapse and shift must vary (because we need to evolve the full set of solution trajectories), yet the lapse and shift cannot vary since they are not canonical variables.  The unavoidable outcome of this analysis is that lapse and shift must be retained in the canonical formalism to have any hope of retaining the full diffeomorphism symmetry.

But now we encounter another problem. If the lapse and shift are canonical variables we must insure that variations of these variables engendered by infinitesimal coordinate transformations are faithfully reproduced as canonical transformations. This is the Legendre transformation projectability requirement.

Consider, for example the variation of the lapse function $N$ under an infinitesimal coordinate transformation $x'^\mu = x^\mu - \epsilon^\mu$. The resulting variation of the lapse is
\beq
\delta N(t) := N'(t) - N(t) = \dot N \epsilon^0 + N \dot \epsilon^0.
\eeq
But  since the Lagrangian does not depend on $\dot N$ it is possible to show that projectible functions in configuration/velocity space may not depend on $\dot N$. Therefore $\epsilon^0$ must depend on $N^{-1}$. Similar observation holds for time derivatives of the shift functions. It is then easy to show that  projectable infinitesimal spacetime coordinate transformations are of the form
\beq
x'^\mu = x^\mu - \delta^\mu_a \xi^a(x, g) - n^\mu \xi^0(x, g),
\eeq
where the arbitrary functions $\xi^\mu$ may depend explicitly on the coordinates $x$ and non-locally on the spatial metric components $g$, and $n^\mu = ( N^{-1}, -N^{-1} N^a )$ is the normal to the hypersurface of fixed coordinate time  $x^0$.  It is important to observe that global translations in time, of the form $x'^0 = x^0 + constant$ are not projectable, i.e., time evolution in phase space cannot be put in correspondence with a Lagrangian symmetry. Since the projectable infinitesimal diffeomorphisms have a compulsory dependence on the lapse $N$ and shift $N^a$ metric functions it is best to view the relevant symmetry group as a group of transformations on the metric and its time derivatives. Indeed, in the generic four-dimensional case, the full group depends non-locally on the spatial components of the metric due to the appearance of their spatial derivatives in the group Lie algebra.

The Dirac Hamiltonian, the generator of time evolution, takes the form 
\beq H_{\lambda} = N^\mu {\cal
H}_{\mu} + \lambda^\mu P_\mu ,  \label{HD} 
\eeq where $\lambda^\mu$ are
arbitrary functions of spacetime coordinates. The $P_{\mu}$ are the vanishing momenta conjugate to the lapse and shift, and the physical phase space is further constrained by the secondary constraints ${\cal H}_\mu = 0$.

The projected infinitesimal symmetry variations are generated in phase space by the generators
\begin{equation} G_{\xi }(t) = P_{\mu} \dot\xi^{\mu} + ({\cal H}_{\mu} +
P_{\alpha''}n^{\nu'}{\cal C}^{\alpha''}_{\mu \nu'})\xi^{\nu}\ , \label{gen}
\end{equation}
where the structure functions are obtained from the
closed Poisson bracket algebra
\begin{equation} \{ {\cal H}_{A},{\cal
H}_{B'} \} = {\cal C}^{C''}_{AB'} {\cal H}_{C''}\ ,
\end{equation} and
where spatial integrations at time $t$ over corresponding repeated
capital indices are assumed hereafter. The generators $G_{\xi
}(t)$ act on phase space through the equal time Poisson
brackets, and map solution trajectories into other solutions. In
this sense, it is assumed that all phase
space variables appearing in (\ref{gen}) are solution
trajectories satisfying the Dirac Hamiltonian equations of motion. The generator $G_{\xi }(t)$ transforms solution trajectories into physically equivalent solution trajectories. Generally only one of these transformations in a given equivalence class is a time translation, namely the one for which $\xi^\mu$ happens to be equal to $n^\mu$. We can generate all the members of an equivalence class by fixing a time foliation of spacetime, i.e., selecting initial lapse and shift functions. The action of the symmetry group then produces the full set of lapses and shifts, i.e., alters the time foliation. Thus although it appears that the full diffeomorphism symmetry is lost in the canonical formalism, this is not the case. 

In fact, it was shown in (II) that it might be possible to transform physically inequivalent solutions to solutions that satisfy common coordinate conditions. In particular, it might be possible to find a global spacetime scalar function of the phase space variables that is a monotonically increasing function of the coordinate time for all physically distinct solutions. The same observation is made for spatial coordinates. Such global spacetime scalar functions are then chosen as intrinsic coordinates.  All solutions in the intrinsic coordinate system are invariants under the action of the induced-diffeomorphism group, as we now demonstrate. Let $y(x)$ represent phase space solution trajectories, and let $S^\mu(y(x))$ represent four spacetime scalar functions, i.e., with the property that under the permissible coordinate transformations $x'(x)$
\beq
S^\mu(y'(x')) = S^\mu(y(x)). \label{scalar}
\eeq
Define intrinsic coordinates through the coordinate transformation 
\beq
X^\mu(x) := S^\mu(y(x))=: a^\mu(x),
\eeq
with inverse
\beq
x(X) = a^{-1}(X).
\eeq
Next suppose that a coordinate transformation is undertaken from $x$ to $x'$, and the transition to intrinsic coordinates is undertaken from the primed coordinate system, i.e., 
\beq
X'^\mu(x') := S^\mu(y'(x')),
\eeq
Then it follows from (\ref{scalar}) that the coordinate transformation from $X^\mu$ to $X'^\mu$ is the identity transformation:
\beq
X'^\mu(X) = S^\mu(y'(x'))= S^\mu(y(x)) = a^\mu (x) = a^\mu(a^{-1}(X)) = X^\mu.
\eeq
In other words, every geometrical object evaluated in the intrinsic coordinate system is invariant under coordinate transformations from $x$ to $x'$. Indeed, this is not only a sufficient but also a necessary condition for invariance. 

We are able to construct invariants that exploit the diffeomorphism-induced symmetry group. It can be shown in general that the following procedures are equivalent:
\begin{itemize}
\item Evaluate dynamical variables in the intrinsic coordinate system $X^\mu (x)$.

\item Impose the gauge conditions $x^\mu = S^\mu (y(x))$ and construct the corresponding Dirac bracket algebra. The idea here is that one and only one member of each equivalence class of solutions is selected with this gauge choice, namely the one for which the coordinate time and spatial dependence of the spacetime scalar functions are the trivial dependencies $S^\mu (y(x)) = 
x^\mu$.  Generally each member will evolve under a different Hamiltonian, i.e., with different functions $\lambda^\mu$. However, all solutions will still obey the second-order Einstein equations.  This gauge choice might not be globally, or even locally, implementable. It might eliminate entire equivalence classes. 

\item Gauge transform phase space variables under the action of the diffeomorphism-induced group to solution trajectories that satisfy the gauge conditions above. The resulting gauge invariant variables satisfy the Dirac bracket algebra.
\end{itemize}

We will illustrate all of these ideas with our cosmological model.

\section{Classical symmetry reduction and Hamiltonian dynamics}
\label{sect2}

We take the  metric to be of the form 
\beq
\ell_P^2 g_{\mu\nu}=\ell_P^2 \left(\begin{array}{cccc}-n^2(t) & 0 & 0 & 0 \\0 & a^2(t) & 0 & 0 \\0 & 0 & b^2(t) & 0 \\0 & 0 & 0 & c^2(t)\end{array}\right),
\eeq
where we have chosen to work in natural Planckian units, where $\ell_P$, $m_P$, and $t_P$ are respectively the Planck length, mass, and time. All of our variables, including the lapse variable $n(t)$, are therefore dimensionless. We also take our space and time coordinates to be dimensionless.  We show first that the substitution of this metric into the general Einstein equations with massless source field $\sqrt{\frac{m_P }{ t_P}} \phi(t)$ yields the same equations of motion obtained by substituting this metric into the Einstein action and then varying the resulting reduced Lagrangian. 

The non-vanishing Christoffel symbols are
\beq
\Gamma^{0}_{00} = \frac{\dot {n}}{n}, \,\,
\Gamma^{0}_{aa} = \frac{\dot {a} a}{n^2}, \,\,
\Gamma^{a}_{0a} = \frac{\dot {a}}{a},
 \eeq
with similar expressions for $a \rightarrow b \rightarrow c$. (In the following we will not repeat this instruction).
The resulting Ricci scalar is
\bea
R&=&g^{\alpha\beta}R_{\alpha\beta}
=-\frac{1}{n^2}\left(-\frac{2\ddot{a}}{a}-\frac{2\ddot{b}}{b}-\frac{2\ddot{c}}{c} \right.  \nonumber \\
&+&\left. \frac{2\dot{n}\dot{a}}{na}+\frac{2\dot{n}\dot{b}}{nb}+\frac{2\dot{n}\dot{c}}{nc}-\frac{2\dot{a}\dot{b}}{ab}-\frac{2\dot{a}\dot{c}}{ac}-\frac{2\dot{b}\dot{c}}{bc}\right). \label{r}
\eea
and the corresponding non-vanishing Einstein tensor components, where $G_{\alpha\beta}:=R_{\alpha\beta}-\frac{1}{2}g_{\alpha\beta}R$,  are
\beq
G_{00}=\frac{\dot{a}\dot{b}}{ab}+\frac{\dot{a}\dot{c}}{ac}+\frac{\dot{b}\dot{c}}{bc}, \label{g00} \eeq
\beq
G_{aa}=-\frac{a^2\ddot{b}}{bn^2}-\frac{a^2\ddot{c}}{cn^2}+\frac{a^2\dot{b}\dot{n}}{bn^3}+\frac{a^2\dot{c}\dot{n}}{cn^3}-\frac{a^2\dot{b}\dot{c}}{bcn^2}. \label{gaa}
\eeq

Next we compute the non-vanishing components of the stress-energy tensor (which is obtained from the matter action $-\frac{\hbar}{2}\int{d^{4}x\sqrt{-g}g^{\alpha\beta}\phi_{,\alpha}\phi_{,\beta}}$). We find
\beq
T_{00}=\frac{\hbar \dot{\phi}^2}{4},  \,\,
T_{aa}
=\frac{\hbar a^2{}\dot{\phi}^2{}}{4 n^2{}} . \label{taa}
\eeq

Our first dynamical equation follows from the conservation law
\beq
T^{\alpha \mu}{}_{; \mu}=0.
\eeq
Only the $\alpha = 0$ component is non-trivial, yielding 
\beq
\frac{1}{\hbar} T^{0\mu}{}_{; \mu}
=\frac{\dot{\phi}\ddot{\phi}}{2 n^4}-\frac{\dot{\phi}^2{}\dot{n}}{2 n^5}
+\left(\frac{ \dot{a}}{a} +\frac{ \dot{b}}{b}+\frac{ \dot{c}}{c} \right)  \frac{\dot{\phi}^2}{2 n^4}=0.
\eeq
Equivalently,
\beq
\ddot{\phi}-\frac{\dot{\phi}{}\dot{n}}{n}+\left(\frac{\dot{a}}{a}+\frac{\dot{b}}{b}+\frac{\dot{c}}{c}\right)\dot{\phi}=0.
\label{lphidd}
\eeq

The non-trivial Einstein's equations $G_{\mu\nu}=\frac{8 \pi}{\hbar} T_{\mu\nu}$ are
\beq
\frac{\dot{a}\dot{b}}{ab}+\frac{\dot{a}\dot{c}}{ac}+\frac{\dot{b}\dot{c}}{bc}=2 \pi \dot{\phi}^2, \label{econ}
\eeq
and
\beq
\frac{\dot{n}}{n}\left(\frac{\dot{b}}{b}+\frac{\dot{c}}{c}\right)-\frac{\ddot{b}}{b}-\frac{\ddot{c}}{c}
-\frac{\dot{b} \dot{c}}{b c}=2 \pi \dot \phi^2.
\label{add}
\eeq

We next check to see whether we obtain the same equations of motion by first substituting our assumed metric into the Hilbert action and then varying this reduced action. The Hilbert action is
\bea
S&=&\frac{\hbar}{8 \pi}\int{d^4 x{}\sqrt{-g}R}-\frac{\hbar}{2}
\int{d^{4}x\sqrt{-g}g^{\alpha\beta}\phi_{,\alpha}\phi_{,\beta}} \\
&=:&S'_{grav}+S_{matter}.
\eea
Referring to (\ref{r}) and using $\sqrt{-g}=nabc$ the gravitational action becomes
\beq
S'_{grav}=\frac{\hbar}{8 \pi}\int{}d^4 x \left( \frac{d}{dt}\left(\frac{2bc\dot{a}}{n}+\frac{2ac\dot{b}}{n}+\frac{2ab\dot{c}}{n}\right)-\frac{2c\dot{a}\dot{b}}{n}- \frac{2b\dot{a}\dot{c}}{n}- \frac{2a\dot{b}\dot{c}}{n}\right).\eeq

So ignoring the total time derivative, the effective gravitational action is
\beq
S_{grav}=-\frac{\hbar}{4 \pi }\int{}d^4 x\left(\frac{\dot{a}\dot{b}c+a\dot{b}\dot{c}+\dot{a}b\dot{c}}{n}\right).
\eeq
Also we find that
\beq
S_{matter}
=\frac{\hbar}{2} \int \, d^4 x \frac{abc\dot{\phi}^2}{n}.
\eeq

In the following we will ignore the spatial integral. (We could impose a spatial periodicity, in which case the Lagrangian would be multiplied by a dimensionless constant.)  The total reduced Lagrangian becomes
\beq
L=-\frac{\hbar }{4 \pi n}\left(\dot{a}\dot{b}c+a\dot{b}\dot{c}+\dot{a}b\dot{c}\right)+\frac{\hbar abc\dot{\phi}^2}{2n}. \label{lag} \eeq
Variation of this Lagrangian does indeed produce the equations of motion (\ref{lphidd}), (\ref{econ}) and (\ref{add}).

Next we carry out the Hamiltonian analysis based on the reduced Lagrangian (\ref{lag}). The canonical Hamiltonian is
\beq
H_{c}=p_a\dot{a}+p_b\dot{b}+p_c\dot{c}+p_\phi\dot{\phi}+\pi\dot{n}-L. \label{hc}
\eeq
The momenta conjugate to $a$, $b$, $c$, $\phi$, and $n$ are, respectively,
\beq
p_a=\frac{\partial L}{\partial\dot{a}}=-\frac{\hbar}{4 \pi n}\left(\dot{b}c+b\dot{c}\right), \label{pa}
\eeq
\beq
p_\phi=\frac{\partial L}{\partial\dot{\phi}}= \frac{\hbar abc \dot{\phi}}{n},
\label{rho}
\eeq
and
\beq
\pi=\frac{\partial L}{\partial\dot{n}}=0.
\eeq
Solving (\ref{pa}) and (\ref{rho}) for the velocities, 
\beq
\dot{a}=\frac{2 \pi n}{\hbar }\left(-\frac{p_b}{c}+\frac{ap_a}{bc}-\frac{p_c}{b}\right),\eeq
 and substituting into (\ref{hc}), we find
\beq
H_{c}= \frac{\pi n}{\hbar }\left(-\frac{2p_{a}p_{b}}{c}-\frac{2p_{a}p_{c}}{b}-\frac{2p_{b}p_{c}}{a}+\frac{a p_a^2}{bc}+\frac{b p_b^2}{ac}+\frac{c p_c^2}{ab}\right)+\frac{np_\phi^2}{2\hbar abc}.\eeq

The Dirac Hamiltonian is
\beq
H_\lambda=H_c+\lambda\pi{}\label{ham},
\eeq
where $\lambda$ is a positive but otherwise arbitrary function of $t$.
The resulting dynamical equations are
\beq
\dot{a}=\frac{\partial H_\lambda}{\partial p_a}=\frac{2 \pi n}{\hbar}\left(-\frac{p_b}{c}+\frac{ap_a}{bc}-\frac{p_c}{b}\right)\label{adot},
\eeq
\beq
\dot{p_a}=-\frac{\partial H_\lambda}{\partial a}=-\frac{\pi n}{\hbar }\left(\frac{p^2_a}{bc}-\frac{bp^2_b}{a^2c}-\frac{cp^2_c}{a^2b}+\frac{2p_bp_c}{a^2}\right)+\frac{\hbar np_\phi^2}{a^2bc}, \label{pdota}
\eeq
\beq
\dot{\phi}=\frac{\partial H_\lambda}{\partial p_\phi}=\frac{np_\phi}{\hbar abc}, \label{phidot}
\eeq
\beq\dot{p_\phi}=-\frac{\partial H_\lambda}{\partial \phi}=0, \label{pphidot} \eeq
\beq
\dot{n}=\frac{\partial H_\lambda}{\partial \pi}=\lambda,
\eeq
and
\beq
\dot{\pi}=-\frac{\partial H_\lambda}{\partial n}=-\frac{\pi}{\hbar}\left(-\frac{2p_ap_b}{c}-\frac{2p_ap_c}{b}-\frac{2p_bp_c}{a}+\frac{a p^2_a}{bc}+\frac{b p^2_b}{ac}+\frac{cp^2_c}{ab}\right)-\frac{p_\phi^2}{2 \hbar abc}.\eeq
Note that since $\pi$ is required to vanish (a primary constraint), preservation of this constraint over time yields a secondary constraint
\beq
H:=\frac{\pi}{\hbar}\left(-\frac{2p_ap_b}{c}-\frac{2p_ap_c}{b}-\frac{2p_bp_c}{a}+\frac{ap^2_a}{bc}+\frac{bp^2_b}{ac}+\frac{cp^2_c}{ab}\right)+\frac{p_\phi^2}{2 \hbar abc}
= 0. \label{con}
\eeq

\section{General solution of the equations of motion}
\label{sect3}

The equations of motion (\ref{lphidd}), (\ref{econ}) and (\ref{add}) can be easily solved by rewriting them in terms of the proper time $\tau$, defined through the relation $d\tau = n(t)dt$.  The general solution is  
\beq
a(t)=a_0 \left(\frac{n_0t+\int^t_0 dt_2 \int^{t_2}_0 dt_1 \lambda(t_1)+\tau_0}{\tau_0}\right)^{k_a}, \label{genA} \eeq
\beq
b(t)=b_0  \left(\frac{n_0t+\int^t_0 dt_2 \int^{t_2}_0 dt_1 \lambda(t_1)+\tau_0}{\tau_0}\right)^{k_b},\label{genB} \eeq
\beq
c(t)=c_0  \left(\frac{n_0t+\int^t_0 dt_2 \int^{t_2}_0 dt_1 \lambda(t_1)+\tau_0}{\tau_0}\right)^{k_c},\label{genC} \eeq
and
\beq
\phi(t)=\phi_0+\sqrt{\frac{1}{4 \pi}}k_\phi \ln{ \left(\frac{n_0t+\int^t_0 dt_2 \int^{t_2}_0 dt_1 \lambda(t_1)+\tau_0}{\tau_0}\right)}. \label{phit} \eeq
where $\tau_0$ is the value of $\tau$ when $t=0$. (Henceforth we let the variable symbol with subscript zero represent the value of the variable at time $t=0$). The equations of motion (\ref{lphidd}), (\ref{econ}) and (\ref{add}) result in the following conditions on the constants $k_a$,  $k_b$, $k_c$, and $k_\phi$:
\beq
k_a+k_b+k_c=1,\label{constraint2} \eeq
and
\beq
k_\phi =\pm \sqrt{1-(k_a^2+k_b^2+k_c^2)}. \label{constraint3} \eeq

Notice that as a consequence of (\ref{constraint2}) 
\beq
\frac{a(t)b(t)c(t)}{\tau} = \frac{a_0 b_0 c_0}{\tau_0}.
\eeq
Thus we will be overcounting the complete set of solutions unless we fix this constant ratio. Let us therefore select 
\beq
\tau_0 =  a_0 b_0 c_0, \label{tau0}.
\eeq

Let us also find the general solutions for the momenta. Substituting the configuration variable solutions (\ref{genB}), (\ref{genC}), and time derivatives into (\ref{pa}) we find
\beq
p_a(t) = -\frac{\hbar (1-k_a)}{4 \pi a(t)} = p_{a \,0} \left(\frac{ a_0}{a(t)} \right), \label{genp}
\eeq
where we recognize that the initial momentum $p_{a \, 0}$ is
\beq
p_{a \, 0} = - \frac{\hbar (1 - k_a)}{4 \pi a_0}. \label{pa0}
\eeq
Similarly, we find using (\ref{rho}) and (\ref{tau0}) that
\beq
p_\phi(t) = \frac{\hbar k_\phi}{\sqrt{4 \pi}}. \label{rho0}
\eeq

\section{Reparameterization invariants} \label{sec4}
\subsection{Construction of reparameterization invariants} \label{sec41}

The procedure presented in \cite{ps04} is to find a function of canonical phase space variables that transforms as a scalar under the full diffeomorphism group. In addition this scalar field must increase monotonically in parameter time for all physically distinct solutions in order for it to define an intrinsic time. Of the many acceptable choices we will fix an intrinsic time $T$ as,
\beq
T  = a(t) = a_0\left(\frac{n_0t+\int^t_0 dt_2 \int^{t_2}_0 dt_1 \lambda(t_1)+  a_0 b_0 c_0}{a_0 b_0 c_0} \right)^{k_a} .\label{intrint}
\eeq
This is to be understood as a time reparameterization from time $t$ to time $T$.  Henceforth we will represent variables expressed in terms of intrinsic time, in addition to the intrinsic time coordinate itself, with capital letters. We construct invariants by carrying out this transformation on our solutions. In the next section we will demonstrate explicitly that the resulting functions of phase space variables are invariant under the induced symmetry group corresponding  to reparameterizations of the form 
 $t'=t-\frac{\xi(t)}{n(t)}$.
 
Let us first find the invariant expansion factor $B (T) = ab(t(T))$ where we recognized that $b$ is a scalar under the transformation  (\ref{intrint}). It is not necessary to invert  (\ref{intrint}) for $t(T)$ since the quantity $\frac{n_0t+\int^t_0 dt_2 \int^{t_2}_0 dt_1 \lambda(t_1)+   a_0 b_0 c_0}{a_0 b_0 c_0}$ appears in $b(t)$ in the general solution (\ref{genB}). Therefore we need only solve (\ref{intrint}) for this quantity:
\beq
\frac{n_0t+\int^t_0 dt_2 \int^{t_2}_0 dt_1 \lambda(t_1)+   a_0 b_0 c_0}{a_0 b_0 c_0}
=  \left(\frac{T}{a_0} \right)^{1/k_a}. \label{ratio}
\eeq
Therefore
\beq
B(T)=b(t(T))= b_0  \left(\frac{T}{a_0} \right)^{k_b/k_a}. \label{BT}
\eeq
Similar expressions are found for  $C(T)$. (Of course, $A(T) = T$.)
Similarly, we find
\beq
\Phi(T) = \phi(t(T)) = \phi_0 + \sqrt{\frac{1}{4 \pi}}\frac{k_\phi}{k_a}  \ln{\left(\frac{T}{a_0} \right)}. \label{PhiT}
\eeq
Also, since $n(t)$ is a scalar density of weight one,
\beq
N(T)=\frac{n\left(t(T)\right)}{\frac{dT}{dt}}. 
\eeq
Referring to (\ref{intrint}) and  (\ref{ratio}),
\bea
\frac{dT}{dt} &=&\frac{k_a}{ b_0 c_0} \left(\frac{n_0t+\int^t_0 dt_2 \int^{t_2}_0 dt_1 \lambda(t_1)+  a_0 b_0 c_0}{a_0 b_0 c_0} \right)^{k_a -1} n(t)  \nonumber \\
&=& \frac{k_a}{ b_0 c_0} \left(\frac{T}{a_0} \right)^{1 -1/k_a} n(t) .
\eea
Therefore
\beq
N(T) =\frac{b_0 c_0}{k_a} \left(\frac{T}{a_0} \right)^{-1+1/k_a}. \label{NT}
\eeq
Finally, since $p_b(t)$ is a scalar,
\beq
P_b(T) = p_b(t(T)) = p_{b\,0}\left( \frac{a_0 }{T} \right)^{k_b/k_a}.  \label{PT}
\eeq

Summing up so far, we are to interpret the expressions (\ref{BT}),  (\ref{PhiT}), (\ref{NT}), and (\ref{PT}) as functions of a time parameter $T$.  The $T$ dependence of our invariants is an explicit dependence, not an implicit dependence as an argument of the remaining dynamical variables. The invariants depend only on the initial values of the remaining canonical phase space variables. Yet it turns out, as we will demonstrate shortly, that these initial values may be replaced by their evolved values at coordinate parameter time $t$. Indeed, any function $\lambda$ may be chosen in the Dirac Hamiltonian (\ref{ham}).  In other words, the dynamical-variable-dependent contribution to our invariants commutes with the Hamiltonian. Since the Hamiltonian contains the same constraints as the symmetry generator (\ref{gxit}), our putative invariants must indeed be invariants. {\it Thus the fact that it is possible to replace the initial values by solution trajectories for arbitrary $\lambda$ is already a demonstration of invariance.} Nevertheless,  we will compute explicitly below the variations of our invariants generated by (\ref{gxit}). 

Indeed, substituting
\beq
a_0 = a(t)\left(\frac{\tau_0}{n_0t+\int^t_0 dt_2 \int^{t_2}_0 dt_1 \lambda(t_1)+  a_0 b_0 c_0}\right)^{k_a},
\eeq
and a similar expression for $b_0$
we find
\beq
B(T)= b(t)  \left(\frac{T}{a(t)} \right)^{k_b/k_a}. \label{Bt}
\eeq
Similar substitutions into all of the invariant variables result in
\beq
\Phi(T) =  \phi(t) + \sqrt{\frac{1}{4 \pi}}\frac{k_\phi}{k_a}  \ln{\left(\frac{T}{a(t)} \right)}, \label{Phit}
\eeq
\beq
N(T) =\frac{b(t) c(t)}{k_a} \left(\frac{T}{a(t)} \right)^{-1+1/k_a}. \label{Nt}
\eeq
and
\beq
P_b(T) = p_b(t)\left( \frac{a(t) }{T} \right)^{k_b/k_a}.  \label{Pt}
\eeq
These expressions will be used in section \ref{sec43} where we will demonstrate directly their invariance under the time reparameterization-induced symmetry group.

\subsection{Finite transformations along gauge orbits} \label{sec42}

Since we have at our disposal the full reparameterization-induced gauge symmetry group, we can alternatively use this group to construct gauge invariants. The idea is that given any solution of the equations of motion we can undertake a finite gauge transformation to the unique solution satisfying the gauge condition expressing the fact that the intrinsic time has been chosen to be proportional to the scalar field.

The generator $G(\xi(t))$ of infinitesimal variations in phase space variables, induced by infinitesimal reparameterizations $t'=t-\frac{\xi(t)}{n(t)}$, is according to (\ref{gen})
\bea
G(\xi(t))&=&\xi(t) H(t) + \dot \xi(t)\pi(t) \nonumber \\
&=&\frac{\xi(t) \pi}{\hbar} \left(-\frac{2 p_a(t)p_b(t)}{c(t)}-\frac{2 p_a(t)p_c(t)}{ b(t)}-\frac{2 p_b(t)p_c(t)}{a(t)}+\frac{a(t)p_a^2(t)}{b(t)c(t)} \right.  \nonumber \\
&+& \left.\frac{b(t) p_b^2(t)}{a(t)c(t)}
+\frac{c(t)p_c^2(t)}{a(t)b(t)}
+ \frac{p_\phi^2(t)}{2 \pi a(t)b(t)c(t)}\right)+\dot \xi(t)\pi(t). \label{gxit}
\eea
It is assumed in this expression that the phase space variables are solutions corresponding to the Hamiltonian (\ref{ham}) where a choice has been made for the function $\lambda(t)$. Given the finite arbitrary function $\xi(t)$ it is easy to verify that the effect of the finite operator
\beq
 V_{\xi}(s,t) =  exp\left(s  \{ - , \,
G_{\xi }(t)\}_{y_{\lambda}(t)} \right) \ . \label{gevolution}
\eeq
on phase space solutions $y_{\lambda}(t)$ is to replace $n_0t+\int^t_0 dt_2 \int^{t_2}_0 dt_1 \lambda(t_1)$ in the solutions corresponding to an initial choice of $\lambda(t)$ in the Hamiltonian by $n_0t+\int^t_0 dt_2 \int^{t_2}_0 dt_1 \lambda(t_1) + s \xi(t)$. Thus we obtain a one-parameter family of gauge transformed solutions of the form
\beq
n_s(t) := V_{\xi}(s,t) n(t) = n(t) + s \dot \xi(t), \label{ns}
\eeq
\beq
a_s(t) = a_0 \left(\frac{n_0t+\int^t_0 dt_2 \int^{t_2}_0 dt_1 \lambda(t_1)+\tau_0 +s \xi(t)}{\tau_0}\right)^{k_a},  \label{as}
\eeq
\beq
\phi_s(t) = \phi_0+\sqrt{\frac{1}{4 \pi}}k_\phi \ln{ \left(\frac{n_0t+\int^t_0 dt_2 \int^{t_2}_0 dt_1 \lambda(t_1)+\tau_0 +s \xi(t)}{\tau_0}\right)}. \label{phis}
 \eeq
 
Our strategy is to find the descriptor $\xi(t, y_{\lambda}(t)$ that transforms $a(t)$ to intrinsic time, and then to employ this descriptor in the finite gauge transformation of all of the phase space variables. The resulting objects are manifestly invariant. Thus we set
 \bea
t &=&a_{s=1}(t)  \nonumber \\
 &=& a_0 \left(\frac{ n_0t+\int^t_0 dt_2 \int^{t_2}_0 dt_1 \lambda(t_1)+ \tau_0 +\xi(t; n(t),a_0, b_0, c_0) }{\tau_0}\right)^{k_a} \label{T}
 \eea
It is important to recognize that the left hand side of this expression is invariant under symmetry variations of all of the canonical variables. Changes in $a(t)$, for example, corresponding to a given $\lambda(t)$, will be compensated by changes in $\xi(t; n(t),a_0, b_0, c_0)$. In other words, the time $t$ will now be an independent parameter.

Solving (\ref{T}) for $\xi$, we find
\bea
&&\xi(t; n(t), a_0, b_0, c_0) \nonumber \\
&=&a_0, b_0, c_0  \left(\frac{t}{a_0}\right)^{1/k_a} - \left(n_0 t +\int^t_0 dt_2 \int^{t_2}_0 dt_1 \lambda(t_1) + a_0 b_0 c_0 \right).
\eea
According to (\ref{ns}), (\ref{as}) and (\ref{phis}) the resulting gauge transformed invariants are precisely those obtained in   (\ref{NT}), (\ref{BT}), and(\ref{PhiT}). 

\subsection{Demonstration of invariance} \label{sec43}

 Although in sections \ref{sec41} and \ref{sec42} we have already given rigorous indirect demonstrations of the gauge invariance of the phase space functions $B(t)$, $P_b(t)$, $\Phi(t) $and $N(t)$ under reparameterization-induced symmetry transformations, we will now check explicitly that they are indeed invariant under the infinitesimal variations generated by (\ref{gxit}) for arbitrary infinitesimal $\xi(t)$. 

We may take as our canonical variables $y(t)$ at time $t$ the solution pairs  $(a(t), p_a(t))$, $(b(t), p_b(t))$, $(c(t), p_c(t))$, and $(\phi(t), p_\phi(t) = p_\phi)$. Thus the variation of the solution trajectory $b(t)$  at time $t$ is
\beq
\delta a(t)=\{a(t),G(\xi(t))\}_{y(t)}=\frac{2 \pi  \xi(t)}{\hbar}\left(-\frac{p_b(t)}{ c(t)}-\frac{p_c(t)}{ b(t)}
+\frac{a(t)p_a(t)}{ b(t)c(t)}\right), \label{deltaa1}
\eeq
with analagous variations for $b(t)$ and $c(t)$. Referring to (\ref{genp}) and respecting the condition (\ref{constraint2}) we find that 
\beq
\delta a(t) =\xi(t)  \frac{k_a}{b(t) c(t)}. \label{deltaa}
\eeq
Also,
\bea
\delta p_a(t)&=&\{p_a(t),G(\xi(t))\}\nonumber \\
&=&-\frac{\pi \xi(t)}{\hbar}\left(\frac{ 2 p_b(t)p_c(t)}{ a^2(t)}+\frac{p^2_a(t)}{b(t)c(t)}- 
  \frac{b(t)p^2_b(t)}{a^2(t)c(t)}\right.  \nonumber \\
  &-& \left.\frac{c(t)p^2_c(t)}{a^2(t)b(t)}-\frac{p_\phi^2}{2 \pi a^2(t)b(t)c(t)}\right)  \nonumber \\
&=& \frac{\hbar \xi(t) (k_a - k_a^2)}{4 \pi a^2(t)b(t)c(t)}, \label{deltapa}
\eea
where in the last line we substituted the general solutions  (\ref{genp})  and took into account the conditions (\ref{constraint2}) and ( \ref{constraint3}). In addition we find
\beq
\delta \phi(t)=\{\phi(t),G(\xi(t))\}=\xi(t)\left(\frac{p_\phi}{\hbar a(t)b(t)c(t)}\right)
= \frac{ \xi(t) k_\phi}{\sqrt{4 \pi}a(t)b(t)c(t)}, \label{deltaphi}
\eeq
and
\beq
\delta p_\phi = 0.
\eeq

Let us now check to see whether our phase space functions $B(t)$, $P_b(t)$, $\Phi(t)$ and $N(t)$are invariant under these infinitesimal variations. Referring to (\ref{Bt}),  (\ref{deltaa}), and  (\ref{constraint2}) we find
\beq
\delta  B(t)  
 = t^{k_b/k_a} \left( \delta b(t) a(t)^{-k_b/k_a} - b(t) \frac{k_b}{k_a} a(t)^{-1-k_b/k_a} \delta a(t) \right) = 0.
\eeq
Similarly, referring to (\ref{Nt}),  (\ref{deltaa}),  and  (\ref{constraint2}), we find
\bea
\delta N(t) &=& \frac{1}{k_a}  t^{-1+ 1/k_a} \left(a(t)^{1-1/k_a} \delta \left( b(t) c(t) \right) \right.
\nonumber \\
&+& \left. \left( b(t) c(t) \right) \left( 1 - \frac{1}{k_a} \right) a(t)^{-1/k_a} \delta a(t) \right) = 0.
\eea
Also, referring to (\ref{Pt}),  (\ref{deltaa}), (\ref{deltapa}) and  (\ref{genp}),
\beq
\delta P_b(t) =  t^{-k_b/k_a} \left( \delta p_(t) a(t)^{k_b/k_a} + \frac{k_b}{k_a} a(t)^{-1+k_b/k_a}p_b(t) \delta a(t) \right) = 0
\eeq
And finally, referring to  (\ref{Phit}),  (\ref{deltaa}) and  (\ref{deltaphi}),
\beq
\delta \Phi(t) =  \delta \phi(t) - \frac{k_\phi}{k_a \sqrt{4 \pi}}\frac{\delta a(t)}{a(t)} = 0.
\eeq

\subsection{Gauge fixing and counting of degrees of freedom} \label{sec44}

As explained in detail in \cite{ps04} the selection of an intrinsic time coordinate is equivalent to the choice of a time-dependent gauge condition. The gauge condition equivalent to our choice (\ref{intrint}) is
\beq
\chi_1 = t - a(t)  = 0.
\eeq
The requirement that this condition be preserved under time evolution determines a second gauge condition; setting $\frac{d\chi_1}{dt} = 0$ we find
\beq
0 = \frac{d\chi_1}{dt}  = \frac{\partial \chi_1}{\partial t} + \{\chi_1, H_\lambda \} = 1 - k_a \frac{n}{b c}.  \label{gauge1}
\eeq
Thus the second gauge condition is
\beq
\chi_2 = n(t) -  \frac{b(c) c(t)}{k_a}  = 0. \label{gauge2}
\eeq
There are no more gauge conditions since the requirement that $ \frac{d\chi_2}{dt} = 0$ merely fixes the function $\lambda$.

We investigate next the number of independent parameters at our disposal to vary the initial values of our phase space variables. Since there are five configuration variables, $a$, $b$, $c$, $n$,  and $\phi$, the unconstrained phase space has ten dimensions. The two constraints $\pi = 0$ and (\ref{con}) leave eight, and the two additional gauge conditions (\ref{gauge1}) and (\ref{gauge2}) should leave us six independent initial parameters in our invariant solutions. 

We note that all five of the initial values $a_0$, $b_0$, $c_0$, $n_0$,  and $\phi_0$ may be freely specified. However, according to (\ref{tau0}) and  (\ref{pa0}), the only variable parameters in $P_{a \, 0}$, $P_{b \, 0}$, and $P_{c \, 0}$ are $k_a$, $k_b$, and $k_c$. But these three parameters are subject to the conditions (\ref{constraint2}) and (\ref{constraint3}). In addition, the momenta conjugate to $\Phi$ and $N$ are fixed by the other free parameters. Thus we do indeed find that the total number of independent initial values is six, corresponding to a dynamical system with three degrees of freedom. We may interpret the six independent parameters as labels of equivalence classes of solutions, where a single equivalence class is obtained by performing reparameterizations on solutions labeled by fixed values of the set.

Our counting of degrees of freedom in this model appears to conflict with the counting first undertaken by Ashtekar and Samuels in the Kasner model, in which there is no source \cite{ashsam91,ponshep98}.  They pointed out that in the case in which the spatial topology is ${ \Re}^3$ the freedom to undertake global spatial diffeomorphisms reduces the phase space degrees of freedom to only one. We will address  in a future publication the issue of homogeneity preserving spatial diffeomorphisms in the presence of sources. We observe that in any case our counting is correct when the spatial topology is that of a three-torus. Our focus in this paper has been on time reparameterizations.

\section{Conclusion} \label{sec5}

As is the case with all generally covariant dynamical systems, the phase space version of the anisotropic cosmological model with a massless scalar material source is not covariant under changes in the time coordinate that depend only on the original coordinate time, i.e., coordinate transformations of the form \beq
t'=t'(t). 
\eeq
Nevertheless, there does exist a diffeomorphism-induced gauge symmetry that is realizeable as a canonical transformation group.  The infinitesimal elements of the diffeomorphism group are  time reparameterizations of the form $t' = t - \frac{\xi(t)}{n(t)}$, where $\xi(t)$ is an arbitrary infinitesimal function of the time parameter $t$, and $n(t)$ is the lapse variable. The dependence on the lapse is required in order to produce variations of the metric and of the scalar field that are projectable under the Legendre map from configuration-velocity space (the tangent bundle) to phase space (the cotangent bundle).  The resulting phase space variations are canonical transformations, generated in the present model by (\ref{gxit}).
The phase space variables in this generator are solutions of the equations of motion corresponding to a choice of the function $\lambda(t)$, hence to an initial choice for the parameter time $t$. We should view $t$ in this context as a parameter label identifying a one-parameter set of canonical variables. The canonical variables at time $t$ generate on the same variables at time $t$ a variation calculated via Poisson brackets of the same variables. The lapse solution $n(t)$ is included among this one-parameter set of canonical variables. It must be included as a phase space variable in order to be able to implement the finite diffeomorphism-induced symmetry group; successive infinitesimal diffeomorphisms depend on the transformed lapse. 

We have employed this symmetry group in this paper to display in a non-trivial general relativistic model the time dependence of the metric and scalar variables that are invariant under diffeomorphism-induced symmetry transformations. The conventional argument against the existence of such variables is that if a variable cannot change its value under an arbitrary diffeomorphism $t'(t)$, then obviously the variable must be constant in time. The problem with this argument is that $t'(t)$ is not realizable as a phase space canonical symmetry group, as pointed out above. 

On the other hand, as noted in the Introduction, many authors have argued that non-trivial time evolution can occur when one employs some subset or combination of dynamical fields as a clock. One then seeks correlations between this dynamical clock and the remaining dynamical variables. Komar and Bergmann\cite{komar58,berg61b} pioneered the use of intrinsic time in a canonical constrained Hamiltonian framework, stressing that intrinsic coordinates must be spacetime scalars. But they did not fully examine the canonical symmetry group. In particular, they did not retain the lapse and shift as canonical variables. They were, however, the first to observe that an enlarged diffeomorphism symmetry group existed, and that it possessed a compulsory dependence on the lapse and shift 
 \cite{berg-kom72}.  The resulting classical diffeomorphism-induced canonical transformation group has recently been studied, also in models in which additional gauge symmetries are present
\cite{pss:1997pr,pss:2000jmp,pss:2000grg,pss:2000pr}. 

As we mentioned in the Introduction, many authors claim that the full diffeomorphism symmetry is lost in the transition to phase space.  Dirac himself was led to ``doubt how fundamental the four dimensional (symmetry) in physics is'' and his remark stimulated Barbour's engagement with timeless dynamical models. \cite{dirac63,barbour99}  Kuchar has always maintained that the full diffeomorphism symmetry group is lost in conventional canonical gravity. We cite  for example his remark in 1971 that the ``canonical formalism necessarily destroys the spacetime covariance of the theory by cutting spacetime into slices'' \cite{kuchar71} and a more recent observation with Kouletsis in an extension of a canonical covariant system that ``Our examination of the status of 
of spacetime diffeomorphisms in canonical description of covariant
systems gives no support to a commonly held belief
that spacetime diffeomorphisms are generated by the Hamiltonian
and momentum constraints''. \cite{kouletsisK02} Kuchar and his collaborators maintain that additional embedding variables must be included in order to recover general covariance in the canonical formulation of general relativity.
\cite{ishamk85a,ishamk85b,kuchart91,kucharr95,bicakk97} 

Confusion over the role of the lapse is evident, for example,  in Marolf's treatment of time-reparameterization invariants in cosmology \cite{marolf95}. Although Marolf recognizes that  the canonically implementable infinitesimal  time reparameterization symmetries are of the form $t' = t - \frac{\xi(t)}{n(t)}$, in posing the condition that variables transform as scalar densities of weight one under this transformation he implicitly requires that $\xi(t)$ is constant \footnote{Given that $\omega$ is a scalar density of weight one, it must transform under the canonically realizeable reparameterization $t' = t - \frac{\xi(t)}{n(t)}$ as $\delta \omega = \frac{d}{dt} \left(\omega \frac{\xi}{n} \right)$. The time derivative of $\xi$ is missing in Marolf's equation (3.10)}. More recently, making the assertion that ``translations in time are gauge symmetries'', Dittrich has employed a time independent lapse in her version of an implementation of the Rovelli construction of complete observables. \cite{drittrich04,drittrich06,rovelli01a}. 
Since the lapse variable is generally not a gauge symmetry parameter, one is achieving a group average only in a very restricted context when integrating over it in the manner first advocated by Rovelli \cite{rovelli98}. (Since spatial diffeomorphisms are realizable as canonical transformations without a dependence on the lapse and shift, it is possible that in contrast  the shift functions might double as group parameters.) Rather, the descriptor $\xi(t)$ plays this role. In fact, in averaging over $n(t)$ one is discarding significant physical information. We discuss in the Appendix the limited sense in which time translation may be considered a gauge symmetry.

As we have stressed throughout this work the full reparameterization covariance is retained only when the lapse is retained as a canonical variable and not viewed as an arbitrary coordinate function (and certainly not as a time-independent function). And the lapse in intrinsic coordinates will generally have a non-vanishing Poisson bracket with other canonical variables.

The classical, flat, homogeneous, anisotropic cosmological model with a massless scalar material source is exactly solvable with arbitrary time parametrization. We have chosen the value of one of the expansion factors as an intrinsic time.  Referring to the general solutions of the equations of motion it is manifest that having fixed a set of six initial parameters as discussed in Section \ref{sec43}, for any value of the scalar field there corresponds unique values of the other phase space variables. The correlation is unique; it does not depend on the time coordinate (and correspondingly, the function $\lambda$) with which one happens to be working. In other words, there is a unique map from the intrinsic time coordinate to the remaining field values. This means in particular that if we were to start with a coordinate time $t$ and then transform to a new coordinate time $t'$ before going to intrinsic coordinates, the transformation $T'(t'(t)) = T(t)$. In other words, the composite transformation from $t$ to $T'$ is the same as the original. Thus the symmetry transformation which transforms solutions in the coordinates $t$ to solutions in coordinates $t'$ can have no effect on the values of variables in the intrinsic coordinate system. This is the physical reason why our invariants, constructed as functions of phase space variables in the original time coordinate $t$, must be invariant under diffeomorphism-induced gauge transformations. 

Apart from our use of the full diffeomorpism symmetry group, we want to stress three essential differences between our method of construction of invariants, and the method employed by Dittrich. \cite{drittrich04,drittrich06} First, \underline{every} dynamical  variable in our construction has an invariant form. The reason is that we are in fact merely transforming, via a dynamical-field-dependent coordinate transformation, to a special (intrinsic) coordinate system. Thus, even for just one choice of intrinsic coordinates, there are many more invariants than there are physically distinct solutions. The proof of invariance for the full set of dynamical variables requires that our intrinsic coordinate functions are true spacetime scalars.  Kuchar and his collaborators have also identified problems that arise when the intrinsic time is a spatial scalar and not a spacetime scalar.  \cite{kuchar92b} That our choice of intrinsic time is a spacetime scalar can be demonstrated directly though the symmetry variation generated by (\ref{gxit}), corresponding to the infinitesimal time reparameterization $t' = t - \epsilon(t)$, where $\epsilon = \xi/n $. We find
\beq
\delta T(t) = \left\{ a(t), \xi(t) H \right\}_{y(t)} = \frac{\xi(t)}{n(t)}  \left\{ a(t), n(t) H \right\}_{y(t)}
= \dot T(t) \epsilon(t),
\eeq
which is the transformation law of a spacetime scalar.

Second, physically distinct solutions are identified by their distinct initial conditions. For any choice of spacetime scalar $T$, there are many more Dirac observables (Dittrich's terminology for invariants) than distinct initial conditions. 

Third, our invariants will always satisfy the original second order Einstein equations. Of course, it will very likely not be possible generally to find global spacetime scalars with the correct coordinate dependence. It might perhaps be possible to patch functions together. The ``complete variables'' considered by Rovelli and Dittrich will generally not satisfy the Einstein equations when the chosen ``clock'' partial variable is not a spacetime scalar. This raises deep conceptual and even philosophical issues. These two essential differences will feature in a forthcoming extension of the intrinsic coordinate construction to the infinite-dimensional cylindrical gravitational wave model. \cite{ps07}

In an eventual quantum theory we must expect that the lapse will exist as an operator and that it will be subject to quantum fluctuations, as was previously pointed out for the free relativistic particle \cite{salisbury03}. Finally, we have stressed that evolution in the intrinsic time is non-trivial; time is not really ``frozen" in general relativity. Though it may turn out that no suitable intrinsically defined time exists in generic general relativity, where it does exist, as in this minisuperspace model, it is incumbent on us to explore its implications in an eventual quantum theory of gravity.
This classical analysis suggests a radical new approach to quantum cosmology, and to quantum gravity in general.  \cite{ss06}.

\section*{Acknowledgements}
We would like to thank Larry Shepley for his review of standard Kasner solution techniques, and thanks  to Larry and  to James Friedrichson for many stimulating discussions, and their critical readings of drafts of this paper.  Thanks also to Don Page for some useful comments.  This work was supported by an Austin College Priddy grant. 

\section*{Appendix}

To the best of our knowledge this cosmological model has not been studied from the point of view of diffeomorphism invariants by either Rovelli or more recently, by Dittrich. However, they have clearly described their general program, and we shall implement it in this appendix so as to compare and contrast our results.

In the language of Rovelli, our intrinsic time $T$ is a ``partial observable''. The objective is then to establish a correlation between the value of this intrinsically determined time and the values of the remaining dynamical variables. However, we encounter significant differences in the implementation of this program. First, neither Rovelli or Dittrich insist that their clock variable must always be a spacetime scalar. Second, we disagree in the implementation. Dittrich in particular considers exclusively evolution in proper time, corresponding to a gauge choice $n = constant$. This leads to inconsistencies, as we shall show.  In addition,  diffeomorphism invariants are then constructed out of constants of the motion. The resulting variables are manifestly invariant under global translation in time. That they are also invariant under the full diffeomorphims-induced group is also true, but this property is neither revealed nor explored by these authors.

Dittrich identifies a phase space function $T(y)$ as a clock partial variable. Letting $f(y)$ represent any other independent phase space function, she constructs a corresponding one-parameter family of invariant phase space functions $F_{[f,T]}(\tau,y)$, where $\tau$ is the parameter. The value of $F_{[f,T]}$ for a fixed value of $\tau$ is by definition the value that 
$f(y)$ assumes when $T(y)$ takes the value $\tau$. This formulation would be equivalent to ours at this stage if in addition $T(y)$ were required to be a spacetime scalar. Indeed, Dittrich's extension to field theory would also be equivalent to our introduction of intrinsic coordinates provided her additional ``clock" variables were also spacetime scalars. They would then be interpreted as intrinsic spatial coordinates, in which case a better terminology might be ``rod" variables. Henceforth we will assume that $T(y)$ is a spacetime scalar, and we limit our remarks to the present cosmological model.

We have shown that given any initial choice of time coordinate $t$ it is possible to perform a canonically-generated symmetry transformation on the phase space function $T(y(t))$ such that, for all $t$, $T(y_{s=1}(t)) = t$. In other words, for this presumably unique gauge $\tau = t$. Dittrich does not recognize the existence of the full gauge symmetry group. Rather, she always works in a proper time gauge in which the lapse function $n$ is a constant, so the arbitrary function $\lambda = 0$ and the Hamiltonian is $n H$, where $H$ is the secondary constraint (\ref{ham}).  The only remaining gauge freedom in this case is global translation in the proper time, and it is generated by $ \xi H$, where $\xi$ is a constant. This is indeed a translation when acting on solutions in this special gauge since $t' = t - \frac{\xi}{n} = t - constant $. Furthermore, the functions $F_{[f,T]}$ are invariant under this gauge transformation. 

Let us now implement the Dittrich program. Inspecting the general solutions (\ref{genA}) -  (\ref{genC}) we observe that a constant lapse results in $a(t) b(t) c(t) = n_0 t $, so let us choose $T = a b c $.  This gauge condition results in $n = 1$.The resulting invariant functions in our notation and that of Dittrich are 
\beq
A(\tau_D) = \frac{a(t)}{\left(a(t) b(t) c(t) \right)^{k_a}} \tau_D = F^a_{\left[f^a,T=abc\right]}(\tau_D, y(t)),
\eeq
where we define $f^a(y):= a$.
Similarly,
\beq
\Phi(\tau_D) = \phi(t) + \sqrt{\frac{1}{4 \pi}} k_\phi \ln \left( \frac{\tau_D}{a(t) b(t) c(t)} \right)
= F^\phi_{\left[f^\phi,T=abc\right]}(\tau_D, y(t)).
\eeq
Dittrich proves that her variables $F$ are invariant under translations in $t$; for example,
\beq
 \frac{a(t + \delta t)}{\left(a(t+ \delta t) b(t+ \delta t) c(t+ \delta t) \right)^{k_a}} =  \frac{a(t)}{\left(a(t) b(t) c(t) \right)^{k_a}}.
 \eeq
We have shown, on the other hand, that $t$ is not restricted to proper time; solution trajectories for any choice of $\lambda(t)$ may used. And furthermore, $ \frac{a_\lambda(t)}{\left(a_\lambda(t) b_\lambda(t) c_\lambda(t) \right)^{k_a}} $ is invariant under arbitrary diffeomorphism-induced symmetry transformations generated by (\ref{gxit}).

We need to stress here that the intrinsic time uniquely fixes the Hamiltonian; both the lapse, and it's time derivative, $\lambda(t)$ are determined by this choice. Therefore any choice other than $T = abc$ will lead to inconsistencies in Dittrich's program. To illustrate, let us pick $\tau = a(t)$ and assume that $t$ is the proper time. Then according to (\ref{genA})
\beq
\tau_D = a(t) = a_0 \left( \frac{t}{\tau_0} \right)^{k_a}.
\eeq
Solving for $ \frac{t}{\tau_0}$ and substituting into the other proper time solutions we recover the invariant solutions  (\ref{BT}) and (\ref{PhiT}),
\beq
B(\tau_D)=b_0  \left(\frac{\tau_D}{a_0} \right)^{k_b/k_a},
\eeq
and
\beq
\Phi(\tau_D) = \phi_0 + \sqrt{\frac{1}{4 \pi}}\frac{k_\phi}{k_a}  \ln \left(\frac{\tau_D}{a_0} \right).
\eeq
But there is one crucial difference. These expressions do not satisfy either the Hamiltonian equations (\ref{adot}) - (\ref{pphidot}), or the Einstein equations (\ref{econ}) - (\ref{add}) for $n = 1$. Rather, $n(\tau_D) =\frac{b_0 c_0}{k_a} \left(\frac{\tau_D}{a_0} \right)^{-1+1/k_a}$ and $\lambda(\tau_D) = \frac{\left(-1+1/k_a\right) b_0 c_0}{k_a} \left(\frac{\tau_D}{a_0} \right)^{-1+1/k_a}$. These inconsistencies do not arise in the intrinsic coordinate approach we have pursued in this work, where furthermore it is recognized that the lapse is always an observable. The Dittrich procedure obscures this important point. Generally, the resulting lapse will have an explicit intrinsic time dependence, as well as a dependence on canonical variables, as is the case with (\ref{Nt}). The latter dependence will prove especially significant in quantum gravity since it will produce quantum fluctuations in proper time.

\end{document}